# High-Performance High-Order Stencil Computation on FPGAs Using OpenCL

Hamid Reza Zohouri, Artur Podobas, Satoshi Matsuoka
Tokyo Institute of Technology, Tokyo, Japan
{zohouri.h.aa@m, podobas.a.aa@m, matsu@is}.titech.ac.jp

*Abstract*—In this paper we evaluate the performance of FPGAs for high-order stencil computation using High-Level Synthesis. We show that despite the higher computation intensity and on-chip memory requirement of such stencils compared to first-order ones, our design technique with combined spatial and temporal blocking remains effective. This allows us to reach similar, or even higher, compute performance compared to first-order stencils. We use an OpenCL-based design that, apart from parameterizing performance knobs, also parameterizes the stencil radius. Furthermore, we show that our performance model exhibits the same accuracy as first-order stencils in predicting the performance of high-order ones. On an Intel Arria 10 GX 1150 device, for 2D and 3D star-shaped stencils, we achieve over 700 and 270 GFLOP/s of compute performance, respectively, up to a stencil radius of four. These results outperform the state-of-the-art YASK framework on a modern Xeon for 2D and 3D stencils, and outperform a modern Xeon Phi for 2D stencils, while achieving competitive performance in 3D. Furthermore, our FPGA design achieves better power efficiency in almost all cases.

*Keywords*—FPGA, OpenCL, High-Order Stencil

## I. INTRODUCTION

The stencil computation pattern is widely used in High-Performance Computing (HPC) for weather prediction, seismic and wave propagation simulation, fluid simulations, image processing and convolutional neural networks. This computation pattern involves traversing a multi-dimensional grid, cell by cell, and calculating a weighted sum based on the value of the cell and its neighbors up to a certain distance that is called the *radius* of the stencil. First-order[1] stencils are regularly used in image processing and convolutional neural networks. However, many scientific applications *require* high-order stencils in their computation. Three out of the nine nominees for the ACM Gordon Bell award in the past two years accelerated applications that involved high-order stencil computation [1, 2, 3], with both of the winners being among them. Compared to first-order (Fig. 1), the difficulty for implementing high-order stencils is two-fold: these stencils are more compute-intensive due to more computation per cell update, and they require more on-chip memory since neighbors are further apart in space.

In the past few years, leveraging High-Level Synthesis (HLS) on FPGAs has gained significant traction since it allows programmers with no knowledge of Hardware Description Languages (HDL) to be able to take advantage of the programmability and better power efficiency of FPGAs [4, 5]. This initiative was further accelerated by the adoption of the OpenCL programming language by both Altera (now part of Intel) [6] and Xilinx [7]. In our previous work [8], we studied the optimization of first-order star-shaped 2D and 3D stencils on FPGAs using Intel FPGA SDK for OpenCL. By combining spatial and temporal blocking, unlike many previous works on FPGAs, we achieved higher performance than the limit imposed by the external memory bandwidth, without restricting input size. Furthermore, using multiple HLS-specific optimizations allowed us to overcome the issues arisen from the added design complexity due to multiple levels of blocking.

In this paper, we build upon our previous work to accelerate high-order stencils on FPGAs. We modify our OpenCL kernel so that apart from performance knobs (block size, vector size, and degree of temporal parallelism), stencil radius is also parameterized. This allows us to synthesize high-order stencils on our FPGA platform just by changing one compilation parameter. Our contributions are as follows:

- We optimize second to fourth-order 2D and 3D stencils using one OpenCL kernel for each, and, to the best of our knowledge, report the highest performance for high-order stencil computation on a single FPGA.

- We show that our combined spatial and temporal blocking technique is capable of handling the higher computation intensity and on-chip memory requirement of high-order stencils, allowing us to achieve similar compute performance to that of first-order stencils.

- We show that our performance model remains correct for high-order stencils, and that we can predict obtained performance with the same accuracy as first-order stencils.

- We show that compared to the state-of-the-art YASK framework [9] on a modern Xeon and Xeon Phi Processor, and a recent GPU implementation [10], our FPGA implementation achieves better performance for 2D stencil computation, and competitive performance for 3D, and better power efficiency in almost all cases.

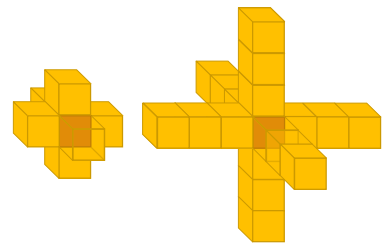

Fig. 1. First-order and third-order star-shaped 3D stencils

---

[1] We consider stencil radius and order to be equal. However, in some scientific publications, stencil order is equal to double the radius and hence, what we call "first-order" is called "second-order."

## II. RELATED WORK

Stencil computation, due to its importance, has been widely studied on different platforms. One of the most prominent works was done by Intel in [11], where they describe the 3.5D blocking technique that allows minimizing redundant computation and external memory accesses for 3D stencil computation. This method involves 2.5D blocking in space (2D blocking and streaming the third dimension), and 1D blocking in time. They provide a performance model to determine best block size, based on the ratio of compute performance to external memory bandwidth of a device (CPU and GPU) and its on-chip memory size. In [12], this method is extended with GPU-specific optimizations, and performance is reported on two newer generations of NVIDIA GPUs. However, neither of these works discuss high-order stencils. Even though there are multiple implementations of application-specific high-order stencils on different hardware, few of them discuss general optimization of such stencils. One of the recent examples on GPUs by Tang et al. uses 2.5D spatial blocking (but no temporal blocking) and an "in-plane" optimization that allows them to achieve better memory alignment and coalescing compared to previous work, at the cost of extra redundant computation for each cell update [10]. They report performance for first to sixth-order stencils.

On Xeon and Xeon Phi processors, one of the most highly-optimized implementations was introduced by Yount [13], which uses a technique called "Vector Folding" that is suitable for wide-vector architectures. This implementation was further optimized and made available to public as the "YASK" framework [9]. We use this framework in our evaluation.

On FPGAs, there is a large body of work on optimizing first-order stencils. The majority of recent examples of such work [14-17] employ temporal blocking to break away from the performance limit imposed by the low external memory bandwidth of modern FPGA boards, but they avoid using spatial blocking. This allows them to achieve linear speed-up with the degree of temporal parallelism, due to lack of halo computation, but comes at the cost of restricting input row size (for 2D) or plane size (for 3D) proportional to the size of FPGA on-chip memory. This restriction will become even more limiting for high-order stencils, due to higher on-chip memory requirement. Among the few implementations of high-order stencil computation on FPGAs, Shafiq et al. [18] implement first to fourth-order 3D star-shaped stencils on a Virtex-4 LX200 FPGA. Their implementation only uses spatial blocking with a cache-like on-chip storage, and input is streamed via DMA to the FPGA. In [19], Fu et al. implement a first-order 3D cubic and a third-order 3D star-shaped stencil using MaxCompiler on two Virtex-5 LX330 FPGAs. Their implementation uses combined spatial and temporal blocking similar to ours.

## III. IMPLEMENTATION

### A. Base Implementation for First-order Stencils

In order to take advantage of the spatial locality of stencil computation and minimize the number of high-latency external memory accesses, we use 1.5D and 2.5D spatial blocking [11], for 2D and 3D stencils, respectively. We also employ temporal blocking to take advantage of the temporal locality of stencil computation by storing intermediate results of multiple iterations (time steps) on-chip, before finally writing them back to external memory. Unlike many previous studies on FPGAs [14-17], combining spatial and temporal blocking allows us to achieve high performance without restricting input size.

We use a single work-item (deep-pipeline) design consisting of a *read*, a *write*, and a *compute* kernel (Fig. 2). The *read* and *write* kernels are connected to external memory to handle memory operations. These kernels are connected to the *compute* kernel via on-chip channels. We defined the *compute* kernel as *autorun*, which allowed us to efficiently replicate this kernel as a 1D array of Processing Elements (PE), each of which operates on a spatial block of a different iteration (time step) in parallel. Data is streamed from the *read* kernel, through the *compute* PEs, and finally written back to external memory by the *write* kernel.

We implement spatial blocking by taking advantage of the shifting pattern of stencil computation, and use shift registers that are implemented using FPGA Block RAMs as on-chip buffers to minimize usage of FPGA on-chip memory. This technique is regularly used for stencil computation on FPGAs [14, 15, 17], but cannot be used on CPUs and GPUs due to lack of hardware support. We also vectorize the computation of each spatial block in the *x* dimension by unrolling our main loop to update multiple consecutive cells in parallel. Temporal blocking is realized by chaining the same number of PEs as the degree of temporal parallelism. To eliminate the need for synchronizing halo data between parallel time steps, we use overlapped blocking. We also employ multiple HLS-specific optimizations to reduce the overhead caused by multiply-nested loops that are required for multiple levels of blocking; specifically:

- **Loop collapsing** to reduce area overhead of storing variable and buffer states in multiply-nested loops
- **Exit condition optimization** by removing the dependency of the loop exit condition on the chain of updates and comparisons on index and block variables, and replacing the exit condition with a single accumulation and comparison on a global index variable
- **Padding** relative to the degree of temporal parallelism to reduce unaligned accesses caused by overlapped blocking that result in memory bandwidth waste

Complete details of our implementation and the performance model we use for parameter tuning are discussed in [8].

### B. Extension for High-order Stencils

To extend out base implementation from [8] for high-order stencil computation, multiple modifications were required:

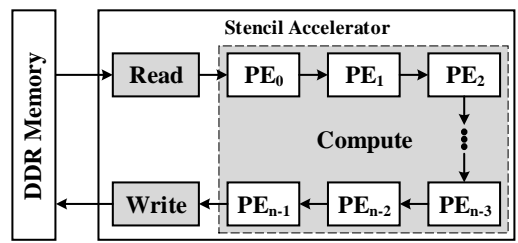

Fig. 2. Design overview, taken from [8]

- Shift register size and addresses for access to different neighboring cells in the shift register were parameterized with respect to stencil radius.
- Comparisons with block and dimension variables were modified to take stencil radius into account. The exit condition for the main loop, which is compared with the global index, was also updated.
- Boundary conditions were modified so that all out-of-bound neighboring cells correctly fall back on the cell that is on the border. Since this could not be efficiently realized using unrolled loops and branches, we created a code generator that generates and inserts the boundary conditions into the base kernel.
- Support for non-square block sizes was added to allow more room for parameter tuning for high-order stencils

Since stencil radius had already been considered in our performance model in [8], no further changes were required in the model to support high-order stencils.

## IV. METHODOLOGY

### A. Benchmarks

We modified the baseline 3D stencil implementation from [12], available at [20], to support high-order stencils. To achieve this, we changed the computation in the main loop to reflect the following equation:

$$f_c^{t+1} = c_c f_c^t + \sum_{i=0}^{rad}(c_w f_{w,i}^t + c_e f_{e,i}^t + c_s f_{s,i}^t + c_n f_{n,i}^t + c_b f_{b,i}^t + c_a f_{a,i}^t) \quad (1)$$

In (1), $f_x^t$ refers to value in position $x$ at time $t$, $c_x$ refers to the coefficient for the value in position $x$, $rad$ is the stencil radius/order, $x \in$ {Center, West, East, South, North, Below, Above}, and the set $\{x,i\}$ refers to the $i^{th}$ neighbor cell in the direction of $x$. In this particular implementation, the coefficient for all the neighbors in a given direction is fixed; however, since we disallow reordering of floating-point operations, the coefficient is not shared and hence, the number of floating-point operations per cell update is equal to $12 \times rad + 1$, with $6 \times rad + 1$ floating-point multiplications (FMUL) and $6 \times rad$ floating-point addition operations (FADD). Optimizing this implementation is equal to optimizing the worst-case scenario where all the coefficients for all of the neighboring cells are different. The 2D version of this kernel uses the same equation, but without the *Above* and *Below* neighbors.

Table I shows the computational characteristics of our benchmarks. The number of bytes per cell update is calculated with the assumption of full spatial reuse (no redundant memory accesses). It is evident from the table that the stencil FLOP to byte ratio increases with its radius, which makes high-order stencils less memory-bound than lower order ones.

### B. Hardware and Software Setup

Our FPGA platform is a Nallatech 385A board with an Arria 10 GX 1150 device and two banks of DDR4 memory operating at 2133 MHz (1066 MHz double data-rate). We used Quartus and Intel OpenCL SDK for OpenCL v16.1.2 for kernel compilations. We avoided newer versions of Quartus (v17.0 and v17.1) since they reliably resulted in lower performance (20-30% lower) and higher area utilization (5-10% more Block RAMs) for the same kernel. For power measurement, we periodically (10 ms interval) read the power sensor that is available on the board using the API provided by the board manufacturer, and average the values during kernel execution.

TABLE I. STENCIL CHARACTERISTICS

| | Radius | FLOP per Cell Update | Byte per Cell Update | $\frac{FLOP}{Byte}$ |
|---|---|---|---|---|
| 2D | 1 | 9 | 8 | 1.125 |
| | 2 | 17 | 8 | 2.125 |
| | 3 | 25 | 8 | 3.125 |
| | 4 | 33 | 8 | 4.125 |
| 3D | 1 | 13 | 8 | 1.625 |
| | 2 | 25 | 8 | 3.125 |
| | 3 | 37 | 8 | 4.625 |
| | 4 | 49 | 8 | 6.125 |

For comparison purposes, we used a 12-core Intel Xeon E5-2650 v4 CPU, and a 64-core Intel Xeon Phi 7210F processor. The Xeon processor is accompanied by quad-channel DDR4 memory operating at 2400 MHz. The Xeon Phi processor was set to operate in flat mode; however, we used *numactl* to set the faster MCDRAM memory as the preferred memory, and made sure the stencil input fits in this memory. All hyperthreads were used on both processors. We implemented our stencils, both in 2D and 3D, using the state-of-the-art YASK framework [9]. Furthermore, we instrumented the framework with power measurement capabilities using the MSR driver [21]. Intel C++ Compiler v2018.1 was used as compiler. It is worth noting that boundary conditions in YASK are different from our implementation. In YASK, the allocated grid is bigger than the input grid so that out-of-bound neighbors can also be read from external memory. This results in extra memory accesses, but allows correct vectorization on grid boundaries. In our implementation, all out-of-bound neighbors fall back on the grid cell that is on the border, instead.

We also compare our results for the 3D stencil with the results reported in [10]. This work uses a different equation in which neighbor cells with the same distance from center share the same coefficient and hence, the FLOP per cell update for their stencil is lower. However, since their implementation is memory-bound for all stencil orders, we assume their reported cell updates per second will be the same if they were not sharing the coefficients (as is done in our implementation). We use the best results from this work that were obtained on an NVIDIA GTX 580. Furthermore, we extrapolate their result for the high-end NVIDIA GPUs of Maxwell and Pascal families, namely GTX 980 Ti and Tesla P100 (PCI-E version), based on the ratio of the theoretical external memory bandwidth of these devices compared to GTX 580. To estimate power usage, we use 75% of the TDP of these GPUs, which closely corresponds to the ratio we *measured* in our previous work [8].

Table II compares the hardware characteristics of the devices we will use for comparison. The peak compute performance is for single-precision floating-point computation. The *FLOP to Byte* column refers to the ratio of compute performance to external memory bandwidth of each device.

TABLE II.     HARDWARE CHARACTERISTICS

| Device | Peak Compute Performance (GFLOP/s) | Peak Memory Bandwidth (GB/s) | TDP (Watt) | Node (nm) | $\frac{FLOP}{Byte}$ | Year |
|---|---|---|---|---|---|---|
| Arria 10 GX 1150 | 1450 | 34.1 | 70 | 20 | 42.522 | 2014 |
| Xeon E5-2650 v4 | 700 | 76.8 | 105 | 14 | 9.115 | 2016 |
| Xeon Phi 7210F | 5325 | 400 | 235 | 14 | 13.313 | 2016 |
| GTX 580 | 1580 | 192.4 | 244 | 40 | 8.212 | 2010 |
| GTX 980 Ti | 6900 | 336.6 | 275 | 28 | 20.499 | 2015 |
| Tesla P100 | 9300 | 720.9 | 250 | 16 | 12.901 | 2016 |

*Without* temporal blocking, computation of a specific stencil will be memory-bound on a given device, if the FLOP to Byte ratio of the stencil is lower than that of the device. Comparing the FLOP to Byte ratios between Table I and Table II, we can conclude that for every stencil order, computation will be memory-bound on all of our hardware. Since the FPGA platform has the highest FLOP to Byte ratio, it is the most *bandwidth-starved* platform and hence, is expected to have the worst computational efficiency. However, due to effectiveness of temporal blocking on FPGAs, we expect to be able to break away from the memory-bound nature of stencil computation on this platform.

### C. Benchmark Settings

For FPGA benchmarks, to minimize redundant computation in the last spatial block and fully show the potential of the device, we set the size of input dimensions to a value that is a multiple of the size of the respective *compute block* dimension. If we denote the size of the compute block in each dimension by $csize_{\{x|y\}}$, we have (from [8]):

$$csize_{\{x|y\}} = bsize_{\{x|y\}} - 2 \times (par_{time} \times rad) \quad (2)$$

In (2), $bsize_{\{x|y\}}$ is the size of the spatial block in each dimension, and $par_{time}$ denotes the degree of temporal parallelism. For 2D stencils, the selected value was between $15500^2$ and $16500^2$, and for 3D stencils, it was between $600^3$ and $750^3$. Exact numbers are reported in Section VI.A. Furthermore, we set the number of iterations to 1000, which resulted in a minimum run time of 3 seconds for 2D, and 11 seconds for 3D stencils. The amount of variation that was observed during FPGA execution was less than 5 ms.

For benchmarks on the Xeon and Xeon Phi processors, we tried multiple different inputs sizes and selected values that gave the best performance. On the Xeon processor, these values were $16384^2$ and $768^3$, for 2D and 3D stencils, respectively, and for the Xeon Phi processor, they were $32768^2$ and $768^3$. All benchmarks used 1000 iterations, for a minimum run time of 53 seconds on the Xeon, and 20 seconds on the Xeon Phi processor.

All of our kernels use single-precision floating-point values. For the FPGA platform, we only measure kernel execution time and ignore data transfer time between host and device. For Xeon and Xeon Phi, we use the timing and performance values reported by YASK, which only include the stencil computation and ignore initialization steps. Our power measurement on YASK starts and ends with the built-in profiler. All experiments are repeated five times, and all values are averaged.

We calculate performance in terms of number of cells updated per seconds (GCell/s) as follows:

$$\frac{run\_time}{num\_input\_grid\_cells \times iteration\_count} \quad (3)$$

Computation performance (GFLOP/s) and throughput (GB/s) values are calculated by multiplying the GCell/s value by the FLOP and byte per cell update values of the stencil, respectively. Redundant computation and memory accesses are *not* taken into account for the reported performance values.

## V. PERFORMANCE TUNING

### A. FPGA

To tune performance on our FPGA platform, we first need to calculate the number of DSPs that are needed to implement each cell update. On the Arria 10 FPGA, each DSP is capable of performing one Fused Multiply and Add (FMA) operation. For each cell update in 2D stencils, $4 \times rad + 1$ multiplications and $4 \times rad$ additions are required. For 3D stencils, $6 \times rad + 1$ multiplications and $6 \times rad$ additions are required. In all cases, each multiplication can be fused with the addition that follows it, except the last one. Hence, $4 \times rad + 1$ and $6 \times rad + 1$ DSPs are required to implement one cell update, for 2D and 3D stencils, respectively. It is worth noting that with shared coefficients, only the number of FMUL operations will be reduced and the number of FADD operations will stay the same. Because of this, DSP utilization will only be reduced by one per cell update, since still one DSP will be required whether the operation is FMA or FADD.

Since the Arria 10 GX 1150 FPGA has 1518 DSPs, we can calculate the total degree of parallelism as:

$$par_{total} = \begin{cases} \left\lfloor \frac{1518}{4 \times rad + 1} \right\rfloor & 2D \\ \left\lfloor \frac{1518}{6 \times rad + 1} \right\rfloor & 3D \end{cases} \quad (4)$$

Then we have:

$$par_{time} \times par_{vec} \leq par_{total} \quad (5)$$

Here, $par_{time}$ and $par_{vec}$ refer to degree of temporal parallelism (equal to numbers of replicated PEs) and vector size, respectively. We restrict $par_{vec}$ to values that are a multiple of two, since the size of ports to memory are limited to such values. We also restrict $par_{time}$ to values that follow (6) to allow best alignment for accesses to external memory (from [8]):

$$(par_{time} \times rad) \bmod 4 = 0 \tag{6}$$

In the next step, based on our previous experience [8], we set a spatial block size (*bsize*) of 4096 for 2D kernels, and a combination of 256×256, 256×128 or 128×128 for 3D stencils, and put all the parameters in our performance model. Then, we extract the top few (usually two) best-performing configurations chosen by our model, and place and route them.

Going from first-order to high-order, we expect the size of the shift register per PE, and consequently, the amount of necessary on-chip memory, to proportionally increase with respect to radius. This is inferred from the following equation for the size of the shift register (from [8]):

$$\begin{cases} 2 \times rad \times bsize_x + par_{vec} & 2D \\ 2 \times rad \times bsize_x \times bsize_y + par_{vec} & 3D \end{cases} \tag{7}$$

Furthermore, based on our earlier discussion, the number of DSPs that are necessary for each cell update will also increase proportional to radius. Intuitively, to optimize parameters for high-order stencils, we can reuse the best configuration for the first-order stencil, but divide the degree of temporal parallelism by radius. We expect this to decrease the number of updated cells per second (GCell/s) proportional to radius, but keep the compute performance (GFLOP/s) nearly the same for all orders since FLOP per cell update increases proportional to radius.

Finally, we use the *flat* compilation flow and avoid Partial Reconfiguration to take full advantage of the available FPGA area and achieve the best timing. We also sweep multiple values of target $f_{max}$ and *seed* to maximize operating frequency.

### B. Xeon and Xeon Phi

The YASK framework includes a built-in performance tuning process that automatically chooses the best block size based on the stencil characteristics and the given hardware. We used the standard flow, which runs this automatic tuner by default and then uses the best parameters for benchmarking. YASK also supports temporal blocking; however, we could not achieve a meaningful performance improvement over what could already be achieved without temporal blocking, regardless of the hardware. Based on the author's recent work [22], temporal blocking in YASK seems to be only effective on Xeon Phi in *cache* mode where a very large input leaks out of the fast MCDRAM to the slower DDR memory, since it allows the more widely-used data to remain in MCDRAM. Since this did not apply to our evaluation, we did not enable temporal blocking.

## VI. RESULTS

### A. FPGA Results

Table III shows our FPGAs results for 2D and 3D stencils from first to fourth-order. Since the OpenCL flow uses the maximum possible $f_{max}$ that meets timing and can be generated by the PLLs on the FPGA, the $f_{max}$ can be an irregular value. The *Estimated Performance* is the value predicted by our performance model [8], normalized for the achieved $f_{max}$.

Our results show that DSP utilization in all cases is equal to what we predicted in Section V.A. Based on the compiler's area reports, for 2D stencils, Block RAM usage per temporal block also increases proportional to stencil radius as we expected. This allowed us to keep the *bsize* the same in all cases. However, rather than dividing the degree of temporal parallelism of the first-order stencil by stencil radius, to get the best configuration for high-order ones, we found other configurations which allowed us to better utilize the DSPs available on the device. On the other hand, with a fixed spatial block size for 3D stencils, the Block RAM utilization per temporal block increased by a factor of 2.5-3 when doubling the stencil radius, which is in contrast to what we expected. This forced us to reduce *bsize* from 256×256 to 256×128 for high-order stencils. We believe that the extra Block RAM usage is either due to some shortcoming in the OpenCL compiler when inferring large shift registers, or some device limitation that requires more Block RAMs than necessary to provide enough ports for all the parallel accesses to the shift register. Here, the best configuration for the high-order 3D stencils could be obtained by dividing the $par_{time}$ value used for the first-order stencil, by the radius of the high-order stencil.

In terms of operating frequency, we see that $f_{max}$ decreases with higher order; however, the critical path of our design only depends on whether the stencil is 2D or 3D, and radius should not affect $f_{max}$. Our tests with smaller parameters ($par_{time}$ and $par_{vec}$) on a Stratix V FPGA showed that the exact same $f_{max}$ could be achieved regardless of the stencil radius. However, for larger parameters on the Arria 10 device, it seems new device-dependent critical paths appear, which further reduce $f_{max}$ as stencil radius increases. As a result, for high-order 3D stencils (second to fourth), we cannot achieve an $f_{max}$ above the operating frequency of the memory controller (266 MHz), which also results in lowered peak memory bandwidth.

In terms of computation throughput (GB/s), we can see that temporal blocking is still effective in every case, allowing us to achieve an effective throughput higher than the available external memory bandwidth. For 2D stencils, we expect temporal blocking to be still effective even for radiuses higher than four; however, for 3D stencils, due to high Block RAM and DSP requirement, fifth and sixth-order stencils will be limited to two parallel temporal blocks, and for higher values, temporal blocking will be unusable. Further accelerating such stencils will only be possible with faster external memory.

TABLE III. FPGA RESULTS

|  | rad | bsize | $par_{vec}$ | $par_{time}$ | Input Size | Estimated Performance (GB/s) | Measured Performance (GB/s\|GFLOP/s\|GCell/s) | $f_{max}$ (MHz) | Logic | Memory (Bits/Blocks) | DSP | Power (Watt) | Model Accuracy |
|---|---|---|---|---|---|---|---|---|---|---|---|---|---|
| 2D | 1 | 4096 | 8 | 36 | 16096×16096 | 780.500 | 673.959\|758.204\|84.245 | 343.76 | 55% | 38%\| 83% | 95% | 72.530 | 86.3% |
|  | 2 | 4096 | 4 | 42 | 15712×15712 | 423.173 | 359.752\|764.473\|44.969 | 322.47 | 64% | 75%\|100% | 100% | 69.611 | 85.0% |
|  | 3 | 4096 | 4 | 28 | 15712×15712 | 264.863 | 225.215\|703.797\|28.152 | 302.75 | 57% | 75%\|100% | 96% | 66.139 | 85.0% |
|  | 4 | 4096 | 4 | 22 | 15680×15680 | 206.061 | 174.381\|719.322\|21.798 | 301.20 | 60% | 78%\|100% | 99% | 68.925 | 84.6% |
| 3D | 1 | 256×256 | 16 | 12 | 696×696×696 | 378.345 | 230.568\|374.673\|28.821 | 286.61 | 60% | 94%\|100% | 89% | 71.628 | 60.9% |
|  | 2 | 256×128 | 16 | 6 | 696×728×696 | 176.713 | 97.035\|303.234\|12.129 | 262.88 | 44% | 73%\| 87% | 83% | 59.664 | 54.9% |
|  | 3 | 256×128 | 16 | 4 | 696×728×696 | 114.667 | 63.737\|294.784\| 7.967 | 255.36 | 44% | 81%\| 99% | 81% | 63.183 | 55.6% |
|  | 4 | 256×128 | 16 | 3 | 696×728×696 | 81.597 | 44.701\|273.794\| 5.588 | 242.77 | 47% | 85%\|100% | 80% | 58.572 | 54.8% |

In terms of compute performance (GFLOP/s), we achieve similar performance for high-order stencils to that of the first-order ones. In fact, for the second-order 2D stencil, we achieve slightly higher GFLOP/s compared to first-order, due to better utilization of the DSPs. However, for other cases, performance slightly decreases due to lower operating frequency. For 3D stencils, even though performance is similar for second to fourth-order, there is a gap between first and second-order. This is due to three reasons: lower DSP utilization, smaller spatial block size, and lower operating frequency. Here, we see another problem of accelerating 3D stencils on FPGAs: due to high FLOP per cell update in high-order 3D stencils, and the restrictions we need to put on our design parameters ($par_{time}$ and $par_{vec}$) to achieve high performance, the number of DSPs used for each PE reaches a few hundred. Because of this, many DSPs are left unused since they cannot be used to accommodate one additional PE.

In terms of updated cells per second (GCell/s), for 2D stencils, performance decreases proportional to the stencil radius. This aligns with what we predicted in Section V.A. For 3D stencils, this relationship is valid for second to fourth-order, but first-order is more than 2x faster than second-order. The reason for this difference is the same as above.

For power usage, we see a gap between the first-order and high-order stencils. The main factor contributing to this difference is the difference in $f_{max}$. The next contributing factor to power usage is area utilization; for example, the third-order 3D stencil uses more power than the second-order one despite lower $f_{max}$, due to higher Block RAM usage.

Finally, with respect to model accuracy, we see that the accuracy is roughly around 85% for 2D stencils, and between 55 to 60% for 3D stencils, which shows that our performance model remains correct for high-order stencils. In practice, this accuracy value reflects the *pipeline efficiency*. As explained in [8], the large difference between 2D and 3D stencils with respect to pipeline efficiency is mainly caused by the larger vectorized accesses used for 3D stencils (higher $par_{vec}$) being split by the memory controller at run time. This results in 40-45% loss of performance for these stencils. We believe the pipeline efficiency is unlikely to improve without major improvements in the memory controller/interface.

### B. Comparison with Other Hardware

Tables IV and V show performance and power efficiency for the 2D and 3D stencils on all applicable hardware. The *Roofline Ratio* column refers to how much of the roofline performance [23] is achieved on each platform assuming full utilization of the theoretical memory bandwidth. The numbers reported in this column effectively show the percentage of the utilized external memory bandwidth, which will be less than 1.00, unless temporal blocking is used. Hachured rows show extrapolated results. Solid green cells show the highest performance and power efficiency for each stencil with each order, *excluding* the extrapolated results. Hachured green cells show these values *including* extrapolated results. Fig. 3 and 4 also show performance for 3D stencil computation on all devices in a more comparable manner. Extrapolated results are hachured.

For 2D stencils, the FPGA achieves the highest performance for first to third-order, and the Xeon Phi for fourth-order. However, the FPGA achieves the best power efficiency in all cases by a clear margin. Despite the highly-optimized implementation in YASK, the Xeon and Xeon Phi devices can only utilize ~50% of their external memory bandwidth (roofline ratio). Furthermore, as explained in Section V.B, temporal blocking proved to be ineffective on these devices. On the other hand, the FPGA can achieve multiple times higher computation throughput than its memory bandwidth, due to the effectiveness of temporal blocking on this platform. For the fourth-order stencil, even with temporal blocking, the achieved computation throughput on the FPGA is lower than the utilized memory bandwidth on the Xeon Phi device and hence, the Xeon Phi achieves better performance. We expect the Xeon Phi to be faster than the Arria 10 FPGA also for stencil orders above four.

TABLE IV. 2D STENCIL PERFORMANCE RESULTS

| | rad | Performance (GFLOP/s) | Performance (GCell/s) | Power Efficiency (GFLOP/s/Watt) | Roofline Ratio |
|---|---|---|---|---|---|
| Arria 10 GX 1150 | 1 | 758.204 | 84.245 | 10.454 | 19.76 |
| | 2 | 764.473 | 44.969 | 10.982 | 10.55 |
| | 3 | 703.797 | 28.152 | 10.641 | 6.60 |
| | 4 | 719.322 | 21.798 | 10.436 | 5.11 |
| Xeon E5-2650 v4 | 1 | 45.306 | 5.034 | 0.521 | 0.52 |
| | 2 | 85.255 | 5.015 | 0.942 | 0.52 |
| | 3 | 124.500 | 4.980 | 1.331 | 0.52 |
| | 4 | 165.231 | 5.007 | 1.737 | 0.52 |
| Xeon Phi 7210F | 1 | 222.804 | 24.756 | 1.000 | 0.50 |
| | 2 | 398.735 | 23.455 | 1.774 | 0.47 |
| | 3 | 592.250 | 23.690 | 2.629 | 0.47 |
| | 4 | 759.198 | 23.006 | 3.369 | 0.46 |

TABLE V. 3D STENCIL PERFORMANCE RESULTS

| | rad | Performance (GFLOP/s) | Performance (GCell/s) | Power Efficiency (GFLOP/s/Watt) | Roofline Ratio |
|---|---|---|---|---|---|
| Arria 10 GX 1150 | 1 | 374.673 | 28.821 | 5.231 | 6.76 |
| | 2 | 303.234 | 12.129 | 5.082 | 2.85 |
| | 3 | 294.784 | 7.967 | 4.666 | 1.87 |
| | 4 | 273.794 | 5.588 | 4.674 | 1.31 |
| Xeon E5-2650 v4 | 1 | 61.282 | 4.714 | 0.686 | 0.49 |
| | 2 | 115.225 | 4.609 | 1.235 | 0.48 |
| | 3 | 151.996 | 4.108 | 1.617 | 0.43 |
| | 4 | 205.751 | 4.199 | 2.069 | 0.44 |
| Xeon Phi 7210F | 1 | 288.990 | 22.230 | 1.279 | 0.44 |
| | 2 | 549.300 | 21.972 | 2.428 | 0.44 |
| | 3 | 788.544 | 21.312 | 3.480 | 0.43 |
| | 4 | 1069.278 | 21.822 | 4.714 | 0.44 |
| GTX 580 | 1 | 224.822 | 17.294 | 1.229 | 0.72 |
| | 2 | 358.725 | 14.349 | 1.960 | 0.60 |
| | 3 | 404.928 | 10.944 | 2.213 | 0.46 |
| | 4 | 453.446 | 9.254 | 2.478 | 0.38 |
| GTX 980 Ti | 1 | 393.322 | 30.256 | 1.907 | 0.72 |
| | 2 | 627.582 | 25.103 | 3.043 | 0.60 |
| | 3 | 708.414 | 19.146 | 3.435 | 0.46 |
| | 4 | 793.295 | 16.190 | 3.846 | 0.38 |
| Tesla P100 | 1 | 842.381 | 64.799 | 4.493 | 0.72 |
| | 2 | 1344.100 | 53.764 | 7.169 | 0.60 |
| | 3 | 1517.217 | 41.006 | 8.092 | 0.46 |
| | 4 | 1699.008 | 34.674 | 9.061 | 0.38 |

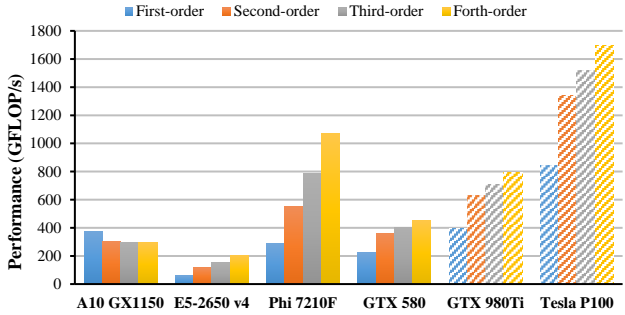

Fig. 3. Performance of 3D stencil computation in GFLOP/s

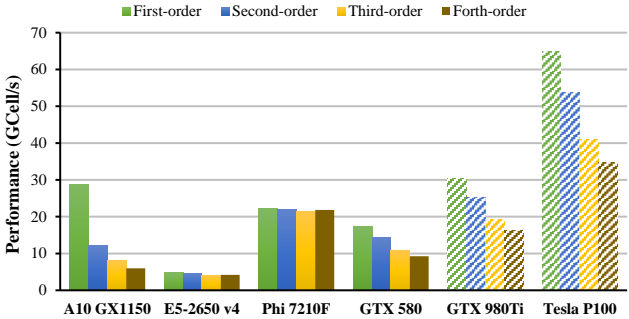

Fig. 4. Performance of 3D stencil computation in GCell/s

For 3D stencils, we also include results from [10] and extrapolated results for newer GPUs. *Excluding* the extrapolated results, The FPGA achieves the highest performance for first-order, while the Xeon Phi and the GTX 580 GPU achieve higher performance for higher orders, with the Xeon Phi achieving the highest; however, the FPGA still achieves the best power efficiency in all orders except four. *Including* the extrapolated results, we expect the state-of-the-art Tesla P100 GPU to be able to achieve the best performance regardless of stencil order, and the highest power efficiency for second-order and higher. In [8], we showed that using the implementation from [12], which also employs temporal blocking, this modern GPU can achieve even higher performance than we estimated here (~50% higher) for the first-order 3D stencil, and better power efficiency than the Arria 10 FPGA. However, this implementation only supports first-order stencils, and the effectiveness of temporal blocking for high-order stencil computation on GPUs is unknown.

Fig. 3 and 4 allow us to see the trend of performance on difference devices with respect to stencil order. On the FPGA, number of cells updated per second (GCell/s) reduces proportional to stencils order, which means the compute performance (GFLOP/s) stays relatively close. On the Xeon and Xeon Phi processors, number of cells updated per second remains similar, which means the compute performance (GFLOP/s) increases proportional to stencil order. On the GPUs, GCell/s decreases with a rate lower than the increase in radius and hence, GFLOP/s increases sub-linearly with stencil order. These performance trends show that the Xeon and Xeon Phi processors, despite being memory-bound in all cases, can utilize a fixed amount of their memory bandwidth regardless of stencil radius. Computation is also memory-bound on the GPUs; however, utilized memory bandwidth decreases as the stencil order is increased. On FPGAs, the trend is different: we can claim the performance we have achieved resembles a *compute-bound* scenario. Even though the level of the achieved compute performance (GFLOP/s) is far from the peak reported in Table II for the Arria 10 device, we emphasize that this peak can only be achieved with full utilization of all DSPs with FMA operations running at the peak operating frequency of the DSPs (~475 MHz) [24]. This level of performance would be near impossible to achieve in real-world designs. For example, for the first-order 3D stencil, the peak performance is ~870 GFLOP/s at the achieved $f_{max}$. Furthermore, only 1344 DSPs are utilized for this stencil, and in one out of each 7 DSPs, the adder in the DSP is not used, which translates to a DSP occupancy rate of ~85%. This way, the achievable peak performance is effectively reduced to ~660 GFLOP/s. Out of this peak performance, close to 40% is lost due to pipeline and memory controller inefficiency, and the remaining difference with the achieved performance is due to redundant computation to support overlapped blocking.

### C. Comparison with Other FPGA Work

With respect to [18], we cannot compare performance in terms of GFLOP/s since coefficients in their stencils are shared and hence, the FLOP per cell update of their stencil is lower than ours; however, comparing the number of cells updated per second, we achieve close to twice their reported performance for fourth-order 3D stencil computation (2.783 vs. 5.588 GCell/s). The results they report are with the assumption that they have 22.24 GB/s of streaming bandwidth, while the system they use only provides 6.4 GB/s. This assumption is not reasonable since streaming bandwidth, whether it is from FPGA external memory or the link between host and the FPGA, will remain the limiting factor in performance of stencil computation for the foreseeable future, and that is why temporal blocking is used. In their case, since temporal blocking is not employed, the roofline of the performance they can achieve in practice is only 0.8 GCell/s.

Compared to [19], since they also share coefficients, we will use the GCell/s metric for performance comparison. They report 1.54 GCell/s for a third-order 3D star-shaped stencil, while we achieve 7.968 GCell/s, which is over 5 times higher. They also estimate that a future FPGA device that is four times larger than a Virtex-6 SX475T FPGA (roughly the size of the modern Virtex Ultrascale+ VU11P device) can achieve close to 5 GCell/s, while we already achieve higher performance.

### VII. CONCLUSION

In this work, we studied the performance and power efficiency of high-order stencil computation on FPGAs using combined spatial and temporal blocking. We showed that even though temporal blocking exhibits limited or no scalability on CPU, GPU and Xeon Phi platforms, good scalability could be achieved on FPGAs even for high-order stencils. This advantage allowed us to scale the compute performance of high-order stencils to close to what could be achieved with first-order stencils. Furthermore, we showed that our performance model remains effective for high-order stencils, allowing us to quickly choose the best performance parameters in every case. Regardless of stencil order, for 2D and 3D star-shaped stencils, we achieved over 700 and 270 GFLOP/s of compute performance, respectively, on an Intel Arria 10 GX 1150 device. This level of performance is enough to compete with other hardware for high-order 2D stencil computation. However, even though temporal blocking is also effective for high-order 3D stencil computation on FPGAs, it is not enough to compete with high-bandwidth devices like Xeon Phi or modern GPUs. We conclude that for high-order 3D stencil computation, having high-bandwidth memory coupled with an efficient memory controller can yield better results *without* temporal blocking, compared to what can be obtained *with* temporal blocking but limited external memory bandwidth. This issue will become even more pronounced for the next-generation Stratix 10 GX 2800 FPGA since the FLOP to byte ratio goes beyond 100 (with 4 banks of DDR4-2400 memory), but the Stratix 10 MX series with HBM memory will likely not suffer from this problem.

ACKNOWLEDGMENT

This work was supported by MEXT, JST-CREST under Grant Number JPMJCR1303, JSPS KAKENHI under Grant Number JP16F16764, the JSPS Postdoctoral fellowship under grant P16764, and performed under the auspices of the Real-world Big-Data Computation Open Innovation Laboratory, Japan. We would like to thank Intel for donating licenses for their FPGA toolchain through their university program.